\documentclass{mem}
\usepackage{natbib}
\usepackage{graphicx}
\begin{document}
\def\teff{$T\rm_{eff }$}
\def\kms{$\mathrm {km s}^{-1}$}

\title{Radiative emission of solar features in Ca II K}

\author{S. Criscuoli\inst{1}, I. Ermolli\inst{1}, J. Fontenla\inst{2}, F. Giorgi\inst{1}, \\
M. Rast\inst{2}, S. Solanki\inst{3}, H. Uitenbroek\inst{4}}

  \offprints{I. Ermolli}

\institute{
INAF Osservatorio Astronomico di Roma, Via di Frascati 33, 00040 Monte Porzio Catone, Roma, Italy
\email{serena.criscuoli@oaroma.inaf.it}
\and
LASP University of Colorado, USA 
\and
MPS, Germany
\and
NSO Sacramento Peak, USA
}

\authorrunning{Criscuoli}

\titlerunning{Radiative emission of solar features in Ca II K}

\abstract{
We investigated the radiative emission of different types of solar features in the spectral range of the Ca II K line. 
 We analyzed full-disk 2k x 2k observations from the PSPT Precision Solar Photometric Telescope. The data were obtained by using 
 three narrow-band interference filters that sample the Ca II K line with different pass bands. Two filters are centered in the line core, 
 the other in the red wing of the line. We measured the intensity and contrast of various solar features, 
 specifically quiet Sun (inter-network), network, enhanced network, plage, and bright plage (facula) regions. Moreover, we compared the 
 results obtained with those derived from the numerical synthesis performed for the three PSPT filters with a widely used radiative code on a set 
 of reference semi-empirical atmosphere models. 
 
}
\maketitle{}

\section{Introduction}

The intensity of the Ca II K resonance line observed with spectrographs and Lyot-type filters has long served as a diagnostic of the 
solar chromosphere and as an indicator of magnetic activity. However, the literature is lacking in long term analysis of the radiative 
properties of solar features at this spectral range. Measurements of large-scale phenomena on Ca II K images provide us with observational 
constraints for modeling the solar atmosphere, the differential rotation and meridional circulation, and the behavior of the solar dynamo. 
Moreover, they give input for improving the calibration and analysis of historical Ca II K observations, which were made at several 
observatories for many years. 

We investigated the radiative emission of different types of solar features in the spectral range of the 
Ca II K line.  We analyzed full-disk 2k x 2k observations from the PSPT Precision Solar Photometric Telescope \citep{coulter1994,ermolli1998,rast2008}. The data were obtained 
by using three narrow-band interference filters that sample the Ca II K line with different pass bands. Two filters are centered in the line 
core, specifically PK-027 and PK-010. The other filter, PKR-011, is centered in the red wing of the line. We measured the intensity and contrast of various solar features, specifically quiet Sun 
(inter-network), network, enhanced network, plage, and bright plage (facula) regions. Moreover, we compared the results obtained with 
those derived from the numerical synthesis performed for the three PSPT filters with the RH code (Uitenbroek 2001, 2002) on a set of reference 
atmosphere models (Fontenla et al. 2007, 2009).

\section{Observations}

We analyzed 157 sets of full-disk observations available on the MLSO PSPT archive for the period June 7th to July 31st 2007. 

All the images were pre-processed to apply the standard instrumental calibration \citep{rast2008}. Then, each image was re-sized 
and aligned, by using linear interpolation, to a common reference grid in which the solar disk size is constant. 

\begin{figure}
\centering{
\includegraphics[width=4.5cm]{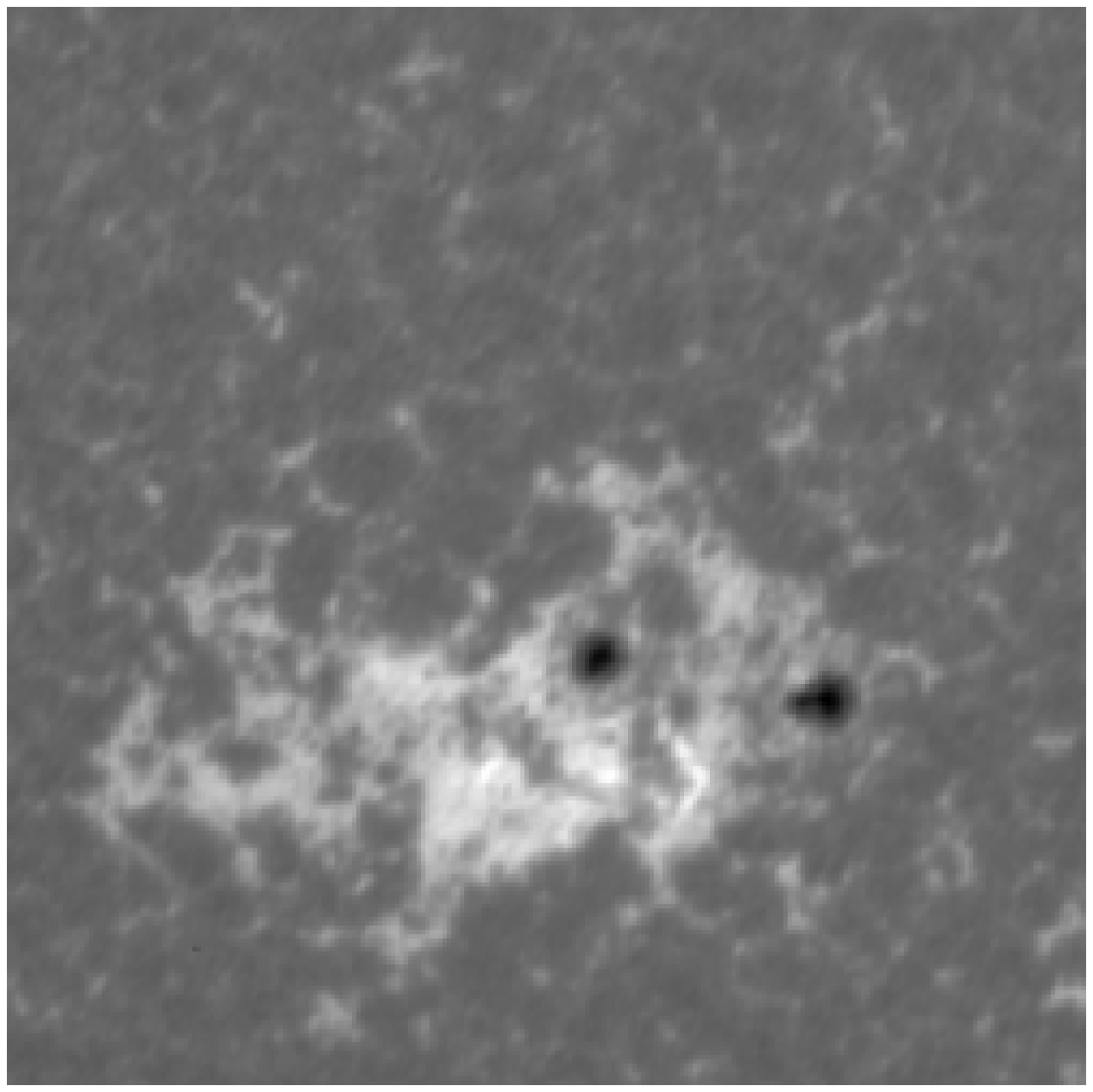}
\includegraphics[width=4.5cm]{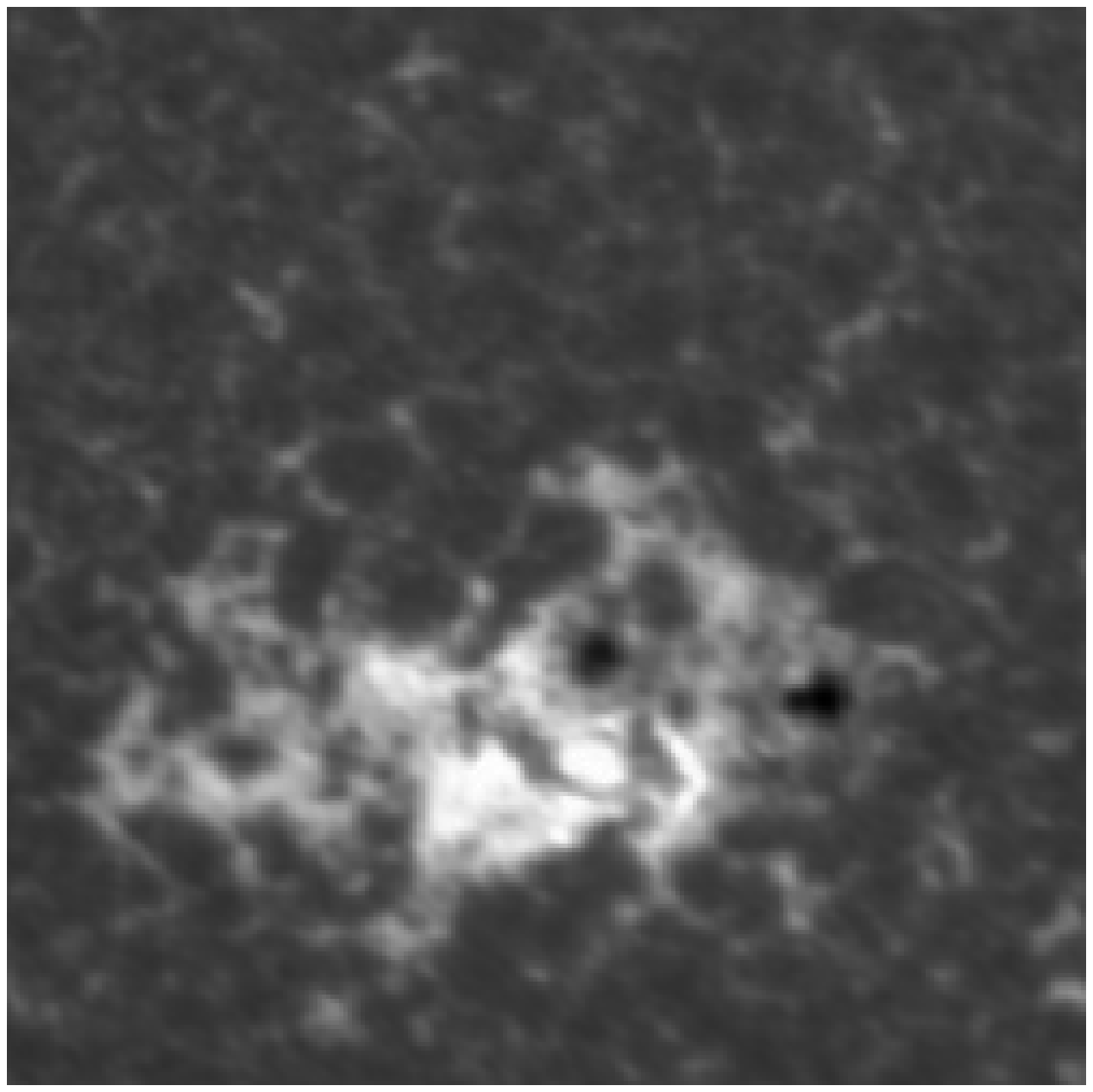}
\includegraphics[width=4.5cm]{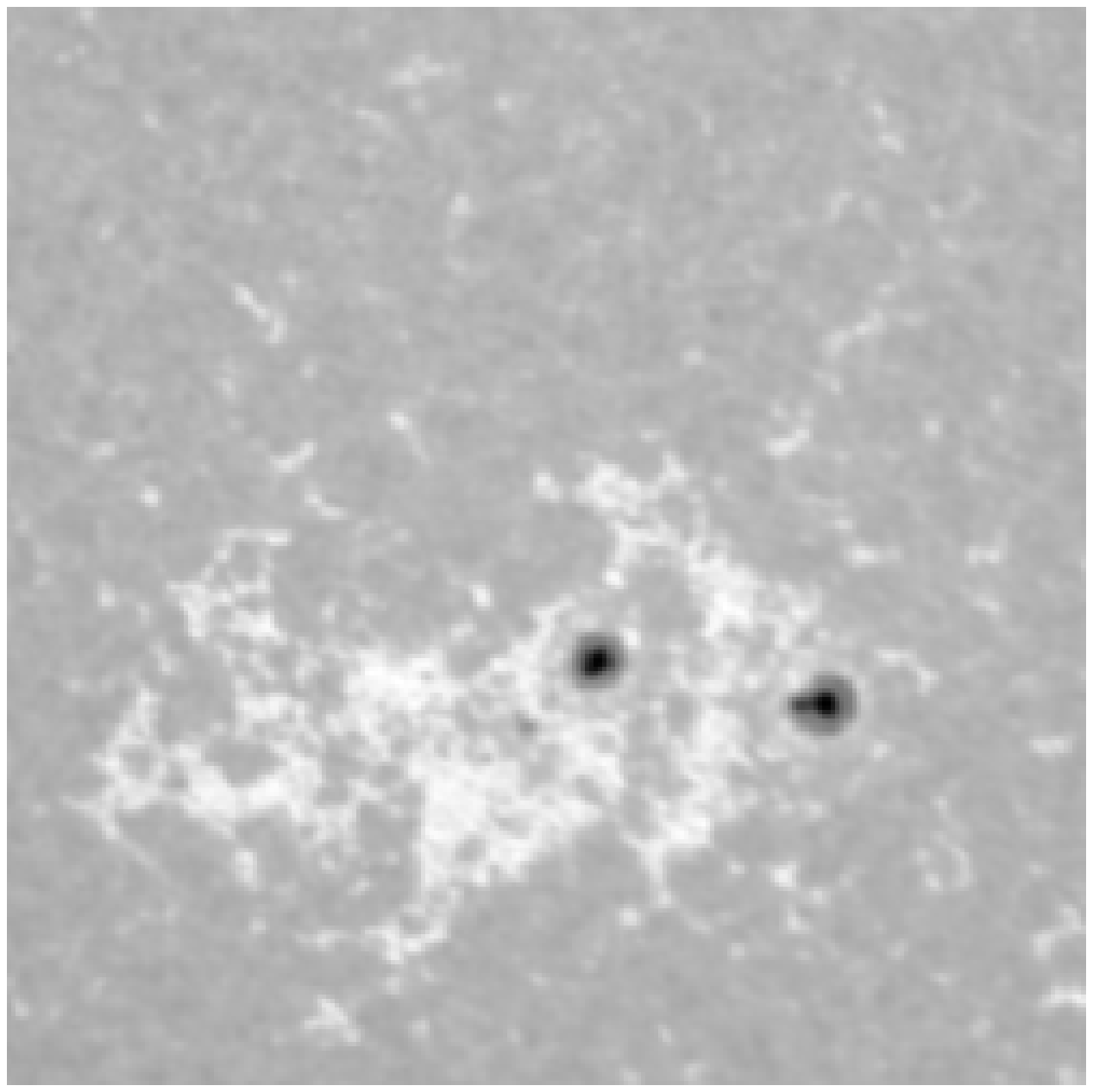}
}
\caption{\footnotesize
Central disk details extracted from the PSPT images obtained with the three filters sampling the Ca II K line 
with different pass bands, specifically PK-027 (top), PK-010 (middle), and PKR-011 (bottom).}
\label{fig1} 
\end{figure}

Figure \ref{fig1} shows central disk details extracted from the PSPT images obtained with the three filters sampling the Ca II K line 
with different pass bands, specifically PK-027 (top), PK-010 (middle), and PKR-011 (bottom). 
Visual inspection reveals the same set of disk features is present in all the images, although their contrast and area look different. 


The 
transmission profiles measured for the three PSPT filters are larger than the 
ones measured for a typical Lyot-type birefringent filter and
do not allow to separate the features of the Ca II K line.

\section{Method}

\begin{figure} 
\centering{
\includegraphics[width=4.5cm]{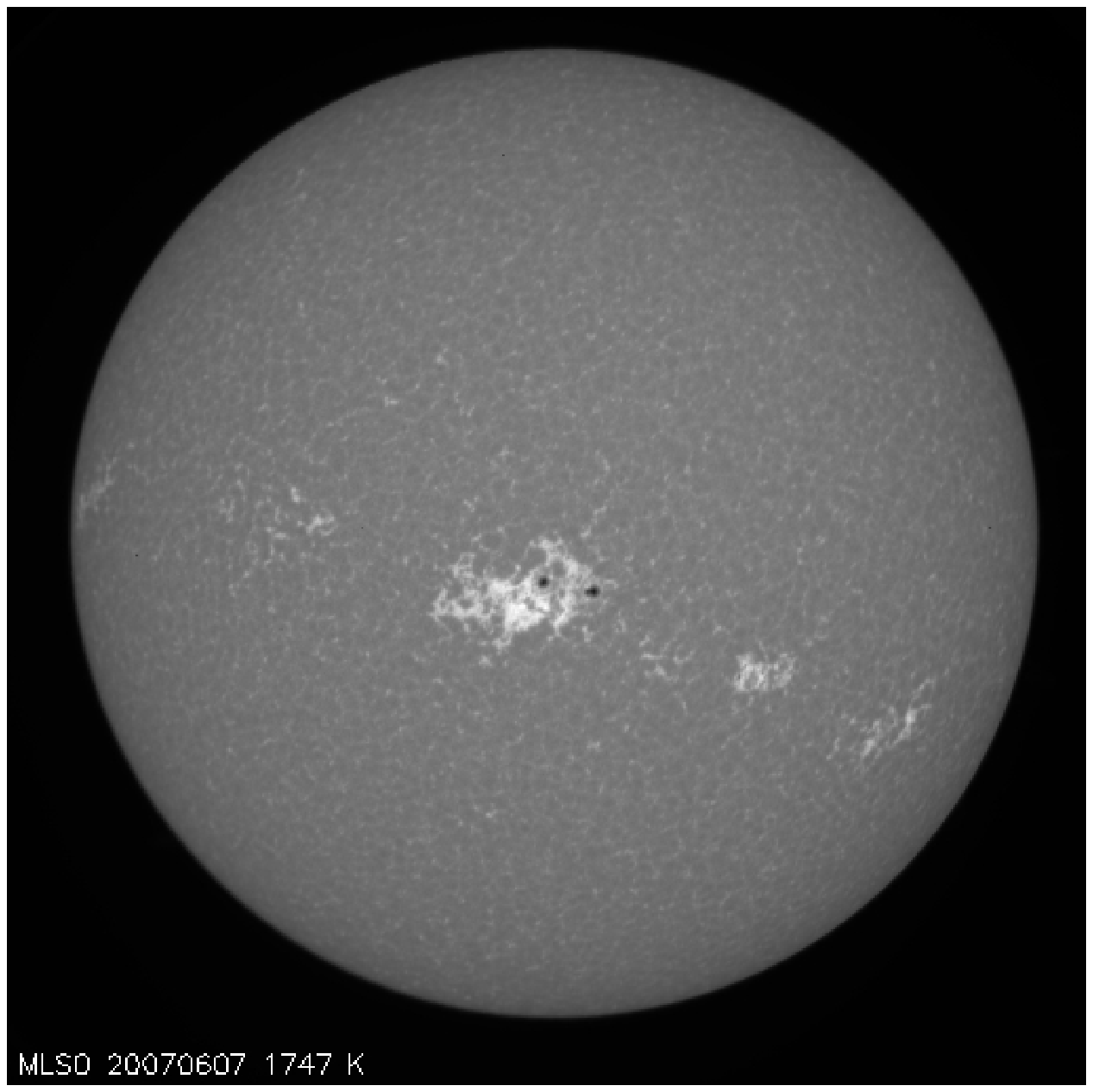}
\includegraphics[width=4.5cm]{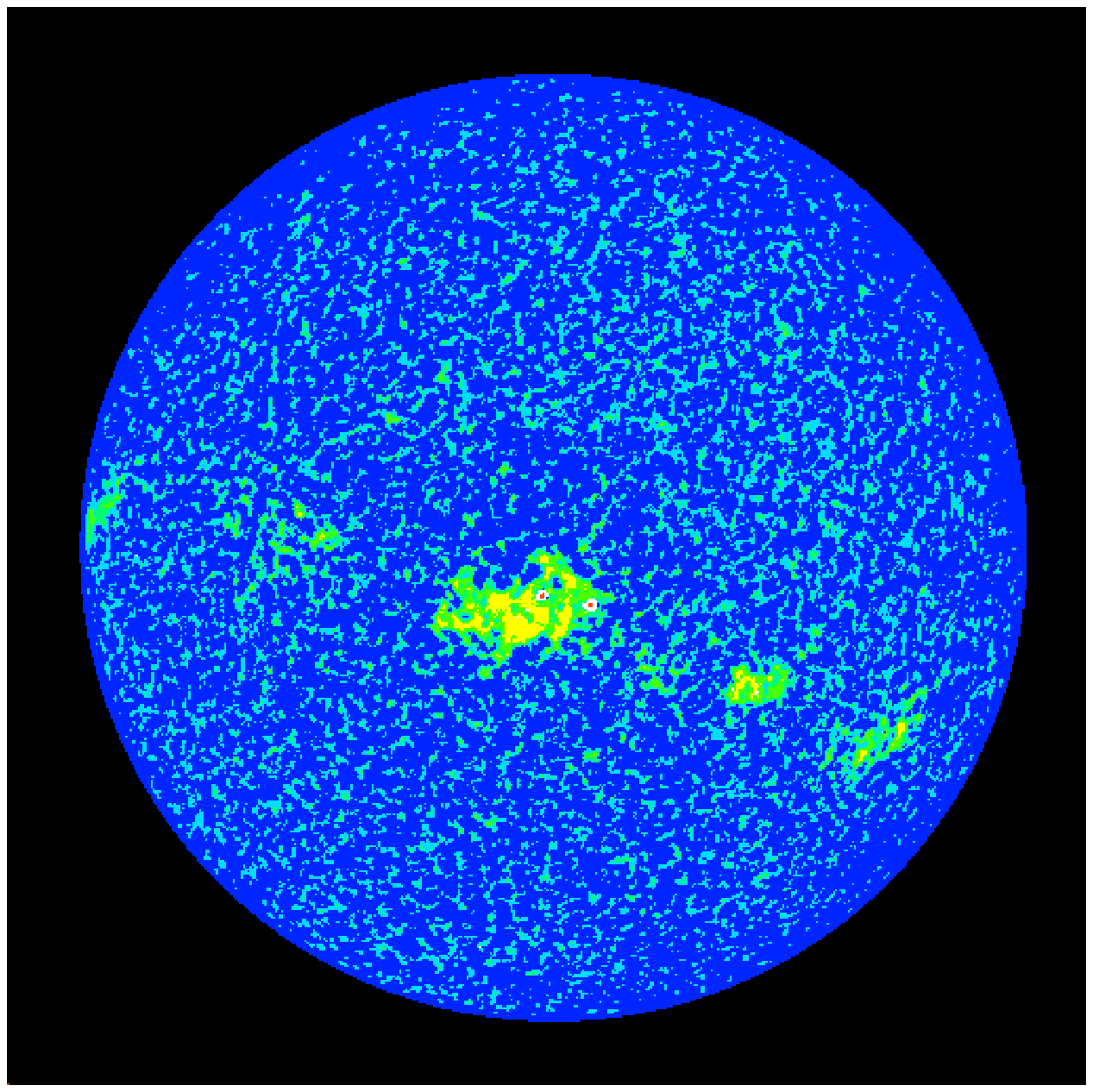} 
}
\caption{\footnotesize
Example of {\it PK-027} (top panel) observation analyzed in this study and of the corresponding mask image (bottom panel). 
This observation was taken on June 7th 2007. The disk features identified on the images 
are show with 
different colors; blue, light blue, green,  and yellow show quiet Sun, network, enhanced network and plage regions; white 
and red show penumbra and umbra regions (visible only blowing up the figure). Black 
shows off-disk regions.}
\label{fig3} 
\end{figure}

We computed synthetic spectra in the range of CaII K with the RH code developed by \citet{uite2001,uite2002}. Our calculations were performed 
assuming nLTE and PRD through the set of atmosphere models presented by \citet{font2007} and \citet{font2009}. 
These models describe the temperature and density vertical stratification of various solar features: quiet Sun (model B), 
network (model D), enhanced network (model F),  plage (model H), and bright plage (model P). 

We used the SRPM image decomposition method  \citep{font2009} to separate the various solar features. 

Figure \ref{fig3} shows an example of the  PK-027 observation and corresponding mask (bottom)
 obtained with the image decomposition. The various disk features are shown in different colors; blue, light blue, green,  and yellow show quiet Sun, network, enhanced network and plage regions.

\begin{figure*}
\centering{\includegraphics[width=12cm]{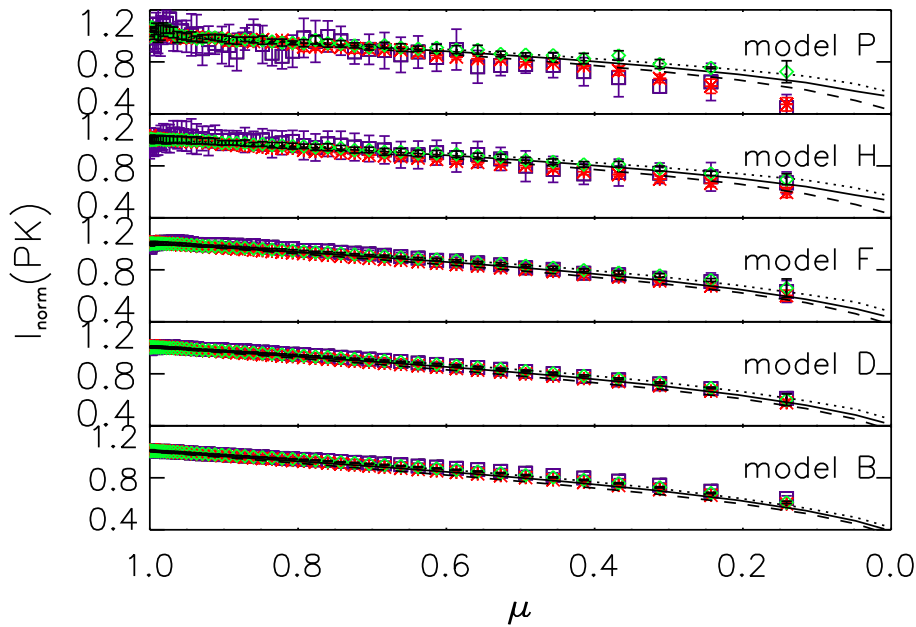}

}
\caption{\footnotesize
Comparison between the CLV of intensity values measured on PSPT images (symbols) 
for the various disk features and those derived from the spectral synthesis (lines) performed for the same filters on the reference atmosphere models. All the values are normalized to the intensity obtained at the disk center. 
The results for the various filters are shown with different colors and line-styles as described in the text.
}
\label{fig4} 
\end{figure*}

\section{Results}

For each atmosphere model, we calculated with RH the Response Functions (RFs) at the disk center for the various PSPT filters  and atmosphere models. These RFs 
indicate that the filters sample quite wide ranges of atmospheric heights, but the radiative signals are dominated by the wings 
of the Ca II K line, that form at heights lower than 500 km.
These RFs also confirm that the narrow band filters sample higher layers than broader ones; the filters in the line core also sample higher 
layers. 

Next, for each filter and disk feature, we analyzed the median and the standard deviation of intensity values measured in 50 equal-area annuli 
around the solar disk center. Figure \ref{fig4}  shows the comparison between the CLV of intensity values measured on PSPT images (symbols) 
for the various disk features and those derived from the spectral synthesis (lines) performed for the same filters on the reference atmosphere models. All the values are normalized to the intensity obtained at the disk center. 
The results for the various filters are shown with different symbols and line-styles; green diamond and solid line, violet square and dotted line, and red asterisk and dashed line show the results obtained for PK-027, PK-010, and PKR-011, respectively. 

We also analyzed the contrast CLV measured on PSPT images for the various disk features.  For each disk position and feature, the contrast was defined as the ratio between the intensity value 
derived for the feature and the value obtained for the quiet Sun. Finally, we compared the results of our measurements with  those 
derived from the spectral synthesis.

\section{Conclusions}
We studied the radiative emission of solar features at the Ca~II~K spectral 
range.  We analyzed moderate resolution PSPT observations taken with interference filters that sample the Ca~II~K range with different pass bands. 
We compared the results of our CLV measurements for various disk features  
with the output of the spectral synthesis performed on the most recent set of 
semi-empirical atmosphere models presented in the literature by Fontenla et al. (2007, 2009). 

We found that the Response Functions derived for the three PSPT filters depend only slightly on the filter band pass and reference 
atmosphere; the most sensitive are the ones derived for the filters with the narrower-band and reference atmosphere of bright plage (model P). 

We also found that the CLV of contrast values measured for the various features observed at the line core with the PK-010 filter are in good
agreement with the outcomes of the spectral 
synthesis. However, the CLV of the contrast obtained for the plage regions (model H) identified on our observations have larger decrease toward the limb than the one derived from the corresponding 
atmosphere model used in this study.

\begin{acknowledgements}
SC acknowledgs the financial support of the NSO. This work has been partially supported by the ASI and INAF, 
within the ASI-ESS and PRIN-INAF-07 contracts, 
respectively. The authors thank ISSI for having hosted meetings of the SSI team. 
\end{acknowledgements}

\bibliographystyle{aa}

\end{document}